\documentclass{article}
\usepackage{spconf,amsmath,graphicx,amssymb,amsfonts}

\usepackage{enumitem}
\usepackage{cite}
\usepackage{amsmath,amssymb,amsfonts}
\usepackage{algorithmic}
\usepackage{graphicx}
\usepackage{adjustbox}
\usepackage{multirow}
\usepackage{textcomp}
\usepackage{color,soul}
\usepackage{xcolor}
\setlist{nosep, leftmargin=14pt}

\usepackage{mwe} % to get dummy images

\makeatletter
\newcommand{\printfnsymbol}[1]{%
  \textsuperscript{\@fnsymbol{#1}}%
}

\title{SWIN-SFTNet : SPATIAL FEATURE EXPANSION AND AGGREGATION USING SWIN TRANSFORMER FOR WHOLE BREAST MICRO-MASS SEGMENTATION}
\name{Sharif Amit Kamran$^{\dagger}$$^{\star}$\thanks{$^{\star}$ Equal Contribution}, Khondker Fariha Hossain$^{\dagger}$$^{\star}$, Alireza Tavakkoli$^{\dagger}$, George Bebis$^{\dagger}$, Sal Baker$^{\mathsection}$}
\address{$^{\dagger}$ Department of Computer Science \& Engineering, University of Nevada, Reno, NV, USA\\
$^{\mathsection}$ School of Medicine, University of Nevada, Reno, NV, USA}
\begin{document}
%\ninept
%
\maketitle
\begin{abstract}
Incorporating various mass shapes and sizes in training deep learning architectures has made breast mass segmentation challenging. Moreover, manual segmentation of masses of irregular shapes is time-consuming and error-prone. Though Deep Neural Network has shown outstanding performance in breast mass segmentation, it fails in segmenting micro-masses. In this paper, we propose a novel U-net-shaped transformer-based architecture, called Swin-SFTNet, that outperforms state-of-the-art architectures in breast mammography-based micro-mass segmentation. Firstly to capture the global context, we designed a novel Spatial Feature Expansion and Aggregation Block(SFEA) that transforms sequential linear patches into a structured spatial feature. Next, we combine it with the local linear features extracted by the swin transformer block to improve overall accuracy. We also incorporate a novel embedding loss that calculates similarities between linear feature embeddings of the encoder and decoder blocks. With this approach, we achieve higher segmentation dice over the state-of-the-art by 3.10\% on CBIS-DDSM, 3.81\% on InBreast, and 3.13\% on CBIS pre-trained model on the InBreast test data set.
\end{abstract}
\begin{keywords}
Breast mass segmentation, Mammogram, Swin Transformer, Deep learning, Medical Imaging
\end{keywords}
\section{Introduction}
\label{sec:intro}

Breast cancer is one of the most dominant cancer types in the world, and Mammography has been acknowledged as a vital tool for the early detection of breast cancer. However, asymmetrical shapes, microcalcifications, and small masses complicate automated breast mass segmentation. Additionally, most computer-aided diagnosis (CAD) systems rely on traditional image-processing-based approaches, which are quite error-prone and require manual intervention. Recently, machine learning and deep learning approaches have outperformed these conventional methods \cite{singh2020breast} and have become a popular technique for such tasks. Nonetheless, most CAD tools are still plagued by manually extracting suspicious regions or segments from low-resolution images, which fail to segment micro masses with accurate contour and high probability. 

Deep Neural Network has shown excellent performance in medical image segmentation. Popular networks like U-Net\cite{ronneberger2015u}, FCN \cite{long2015fully}, AUNet\cite{sun2020aunet}, ARF-Net \cite{xu2022arf} demonstrated outstanding outcomes for breast mass segmentation from both mammography images. These networks implemented diverse methods like generating multi-scale feature maps, attention-guided dense upsampling, and additive channel attention to learn robust feature maps to segment tumors of different sizes with more than 85\%+ dice scores. However, the dice score of these systems falls to 5-15\%  when applied to images with micro-masses. 

One reason for the failure of CNN-based approaches on micro-masses is they overtly focus on global semantic information. And to eliminate similar problems, Vision-transformer (ViT) \cite{dosovitskiy2020image} was proposed to prioritize local patch-level information. Taking 2D image patches with positional embeddings as input, Vision Transformers has outperformed most medical imaging downstream tasks \cite{cao2021swin,kamran2021vtgan,hatamizadeh2022unetr}. Recently, Swin-UNet has achieved phenomenal results in organ segmentation like Gallbladder, Spleen, Liver, etc. Although Swin-UNet can capture local information correctly for precise boundary segmentation of organs, the organ is unique in shape and does not contain similar-looking artifacts. One of the primary problems of segmenting micro-masses in the breast is that surrounding fatty tissues can throw off the segmentation boundary of the model and might raise the false-positive rate as well. To address the above issues, we propose a novel transformer network named Swin Spatial Feature Transformer Network (Swin-SFTNet) and a novel embedding similarity loss to achieve a segmentation dice improvement over the state-of-the-art by 3.10\%, 3.81\%, and 3.13\% on CBIS-DDSM \cite{lee2017curated}, InBreast \cite{moreira2012inbreast}, and CBIS pretrained on InBreast dataset respectively. Our main contributions are: (1) Employing a Swin-Transformer as a basic building block to create Swin-SFTNet to incorporate spatial global and sequential local context information in a multi-scale feature fusion configuration. (2) Designing a novel Spatial Feature Expansion and Aggregation Block to convert sequential linear patches into structured spatial features for capturing global context information for better micro-mass segmentation and (3) Utilizing a novel embedding loss that calculates similarities between features of the encoder and decoder blocks.

\section{Methodology}
\label{sec:methodology}

\subsection{Overall Architecture}
\begin{figure}[!t]
    \centering
    \includegraphics[width=\columnwidth]{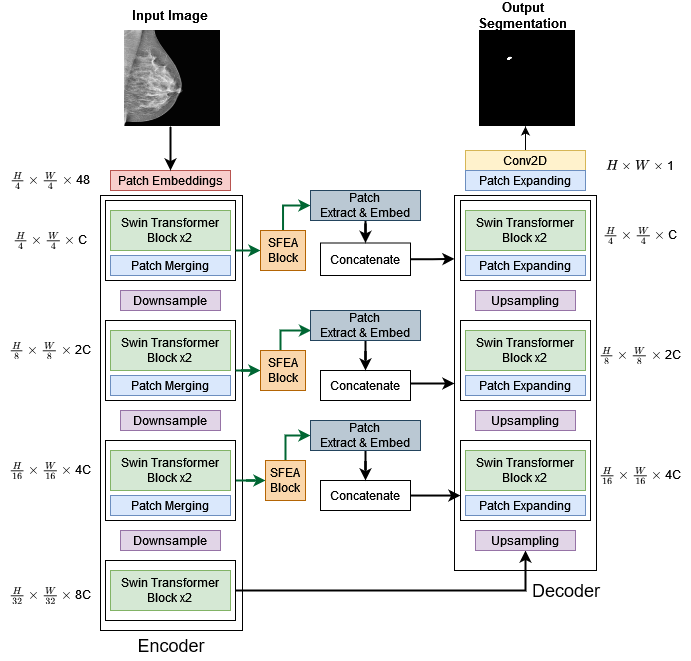}
    \caption{An overview of the proposed Swin-SFTNet consisting of Swin-transformer, Patch Merging, Patch Expanding, Patch Embedding and Spatial Feature Expansion \& Aggregation Blocks.}
    \label{fig1}
\end{figure}

The overall architecture of our proposed Swin-SFTNet is illustrated in Fig.~\ref{fig1}. Swin-SFTNet incorporates an encoder, a decoder, three skip connections between the encoder and decoder, and three parallel SFEA blocks followed by patch extract and patch embedding layer before concatenating with the output feature map. Our architecture is an enhanced version of Swin-UNet \cite{cao2021swin}, a UNet-like auto-encoder that replaces Swin-Transformer blocks \cite{liu2021swin} with regular convolution layers. We first transform the breast mammography grayscale images into RGB, providing the model with learning essential features. We utilize a patch-embedding layer to transform the input into non-overlapping patches of size $4\times 4$. So for three RGB channels, we get to $4 \times 4 \times 3$ = 48 depth dimension. Next, we utilize a dense layer to project feature dimension into C arbitrary dimension. Following this layer, we have our encoder blocks, each consisting of two successive swin-transformer blocks and a patch-merging layer. We explain the swin-transformer block in Subsection \ref{subsec:stb}. We repeat the encoder blocks three times to downsample the feature dimensions from $\frac{H}{4} \times \frac{W}{4} \times C$ to $\frac{H}{8} \times \frac{W}{8} \times 2C$, $\frac{H}{16} \times \frac{W}{16} \times 4C$ and $\frac{H}{32} \times \frac{W}{32} \times 8C$ successively.To conclude the encoder, we utilize two swin-transformer blocks after the last patch-merging layer. 

Similar to the encoder, we design a symmetric decoder composed of multiple Swin Transformer blocks and patch expanding layer. Each decoder black is concatenated with the skip-connection features from the encoder with the same spatial dimension. As a result, we avoid any loss of spatial information due to successive downsampling. In contrast to the patch merging layer, the patch expanding layer reshapes the feature maps with $2\times$ up-sampling of spatial dimension. Additionally, it utilizes convolution to halve the depth dimension. We repeat the decoder blocks three times to upsample the feature dimensions from $\frac{H}{32} \times \frac{W}{32} \times 8C$ to $\frac{H}{16} \times \frac{W}{16} \times 4C$, $\frac{H}{8} \times \frac{W}{8} \times 2C$ and $\frac{H}{4} \times \frac{W}{4} \times C$ successively. The last patch-expanding layer is incorporated to perform 4× up-sampling to restore the resolution of the feature maps to the resolution $H  \times W \times 4C$. We also concatenation operation through the skip-connection features from SFEA and each decoder block's outputs. We explain our proposed Spatial feature aggregation and Expansion block in Subsection \ref{subsec:sfea}. At last we apply a 2D convolution to get the output feature dimension $H  \times W \times 1$ for binary mass segmentation. Here, $H=256$, $W=256$ and $C=128$.

\subsection{Swin-Transformer Block}
\label{subsec:stb}
Traditional window-based multi-head self-attention (W-MSA) proposed in Vision Transformer (ViT) \cite{dosovitskiy2020image} utilizes a single low-resolution window for building feature-map and has quadratic computation complexity. In contrast, the Swin Transformer block incorporates shifted windows multi-head self-attention (SW-MSA), which builds hierarchical local feature maps and has linear computation time. Swin transformer block can be described in the following  Eq.~\ref{eq1} and Eq.~\ref{eq2}.

\begin{equation}
    \begin{split}
    x^{l} &= W{\text -}MSA(\phi(x^{l-1}))\\
    x^{l} &= \delta(\phi(x^{l})) + x^{l}) 
    \end{split}
    \label{eq1}
\end{equation}
\begin{equation}
    \begin{split}
    x^{l+1} &= SW{\text -}MSA(\phi(x^{l}))\\
    x^{l+1} &= \delta(\phi(x^{l+1})) + x^{l+1}) 
    \end{split}
    \label{eq2}
\end{equation}
    
In Eq.~\ref{eq1}, we illustrate the first sub-block of swin transformer consisting of LayerNorm ($\phi$) layer, multi-head self attention module (W-MSA), residual connection (+) and 2-layer MLP with GELU non-linearity ($\delta$). In similar way Eq.~\ref{eq2} illustrates the second sub-block of swin transformer consisting of LayerNorm ($\phi$) layer, shifted window multi-head self attention module (SW-MSA), residual skip-connection (+) and  MLP with GELU activation ($\delta$). Additionally, $l$ notifies layer number and $x$ is the feature-map.

\begin{figure}[!t]
    \centering
    \includegraphics[width=\columnwidth]{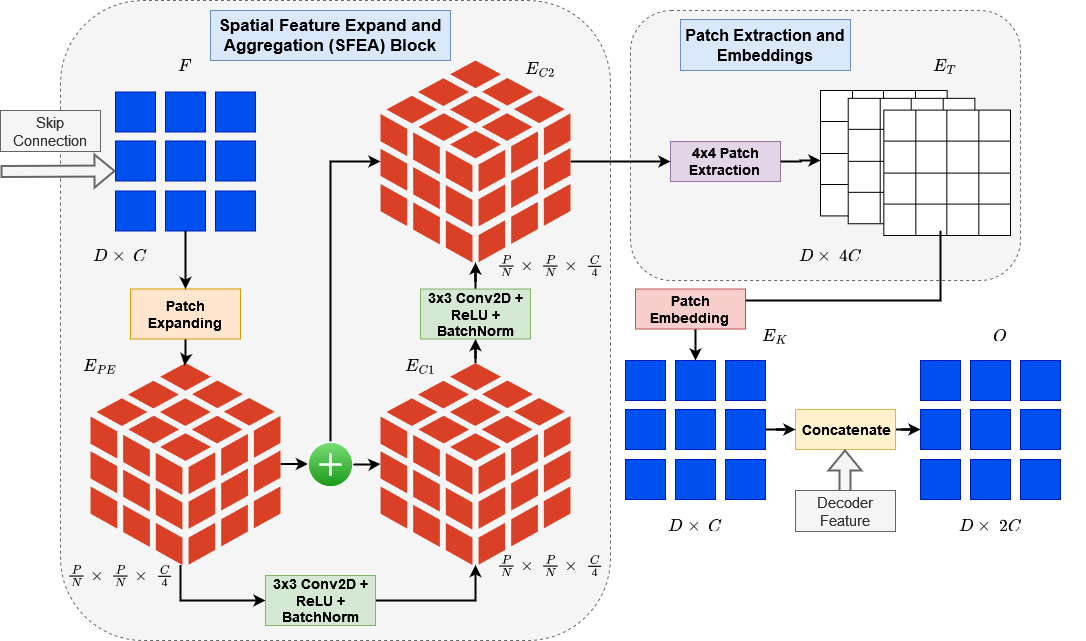}
    \caption{Visualization of the proposed SFEA Block}
    \label{fig2}
\end{figure}

\subsection{Spatial Feature Expansion and Aggregation}
\label{subsec:sfea}
Although the multi-head self-attention module can capture local contextual information to understand inherent feature representations, consecutive patch merging and expanding layers can degrade the overall global context of the task. The existing skip connection concatenation cannot solve this problem as they apply dense layers on sequential patches to create linear projections. To create spatial projections of learnable features, we propose the Spatial Feature Expansion and Aggregation block illustrated in Fig.~\ref{fig2}. We start with the top-most skip connection that comes out of the first encoder layer and has a feature output of $F\in\mathbb{R}^{D \times C}$, where $D=4096$ and $C=128$. We apply a patch expanding layer with patch-size $4\times4$ which gives us the feature output $E_{PE}\in\mathbb{R}^{\frac{P}{N}\times \frac{P}{N}\times \frac{C}{4}}$. Here, spatial dimension $P=256$ and $N=1$, so the resultant spatial dimension becomes $256 \times 256 \times 32$. In a similar manner, from the 2nd and 3rd skip connections with $1024 \times 256$ and $256 \times 512$ dimensional feature input we can get $128 \times 128 \times 64$ and $64 \times 64 \times 128$ feature outputs with $[N_{2},N_{3}]=[2,4]$. Next, we apply a $3\times 3$ 2D Convolution, ReLU activation, and Batch-Normalization operation followed by element-wise addition of features from $E_{PE}$ to get output feature $E_{C1}\in\mathbb{R}^{\frac{P}{N}\times \frac{P}{N}\times \frac{C}{4}}$. In a similar manner, we apply another same 2D Convolution Block on $E_{C1}$ to get feature output and add element-wise features from $E_{PE}$ to get final output $E_{C2}\in\mathbb{R}^{\frac{P}{N}\times \frac{P}{N}\times \frac{C}{4}}$. These two convolution operation helps with extracting global spatial context information that we further combine with our decoder's local patch-level information. Following this operation, we utilize the  $4\times 4$ Patch-extraction operation to convert it the feature into 2D sequence feature output, $E_{T}\in\mathbb{R}^{D\times 4C}$. After that, we use patch-embedding layer to make the feature dimension same as the decoder's paired output, so the output feature map becomes $E_{K}\in\mathbb{R}^{D\times C}$. Next, we concatenate the feature from the decoder's patch expanding layer with the $E_{K}$. We do this for all of our skip connections, so the output feature map becomes $O\in\mathbb{R}^{D \times 2C}$. Here, we use three different values for $C=[128,256,512]$, for the three skip connections. 

\subsection{Objective Function and Embedding Similarity Loss}
For binary output of background and masses we use binary cross-entropy loss given in Eq.~\ref{eq3}. We also use Dice-coefficient loss given in Eq.~\ref{eq4} for better segmentation output. For dice-coefficient we use $\varepsilon=1.0$ in numerator and denominator for addressing the division by zero. Here, $\mathbb{E}$ symbolizes  expected values given, $p$ (prediction) and $y$ (ground-truth).
\begin{equation}
    \mathcal{L}_{bce} =\mathbb{E}_{p,y} \big[ \frac{1}{N}\sum_{i=1}^{N}-(y_{i}*log(p_{i})+(1-y_{i})*log(1-p_{i})) \big]
    \label{eq3}
\end{equation}

\begin{equation}
     \mathcal{L}_{dsc} =\mathbb{E}_{p,y} \big[ 1-\frac{2\sum_{i=1}^{N}p_{i}y_{i}+\varepsilon}{\sum_{i=1}^{N}p_{i}+\sum_{i=1}^{N}y_{i}+\varepsilon}\big]
    \label{eq4}
\end{equation}

 Finally, the embedding feature loss is calculated by obtaining positional and patch features from the transformer encoder layers $E$ and decoder layers $D$ by inserting the image, as shown in Eq.~\ref{eq5}. Here, $Q$ stand for the number of features extracted from the embedding layers of the transformer-encoder.

\begin{equation}
 \mathcal{L}_{emb} = \mathbb{E}_{x,y}\sum_{i=1}^{k}\frac{1}{Q}\parallel E_{em}^{i}(x)-D_{em}^{i}(x)\parallel
    \label{eq5}
\end{equation}

We combine Eq.~\ref{eq3}, \ref{eq4}, and \ref{eq5} to configure our ultimate cost function as provided in Eq.~\ref{eq6}. Here, $\lambda$ is the weight for each loss.

\begin{equation}
     \mathcal{L} = \lambda_{dice}*\mathcal{L}_{dsc} + \lambda_{bce}*\mathcal{L}_{bce}  + \lambda_{emb}\mathcal{L}_{emb} 
    \label{eq6}
\end{equation}

\begin{figure*}[!tp]
    \centering
    \includegraphics[width=0.8\textwidth,height=5cm]{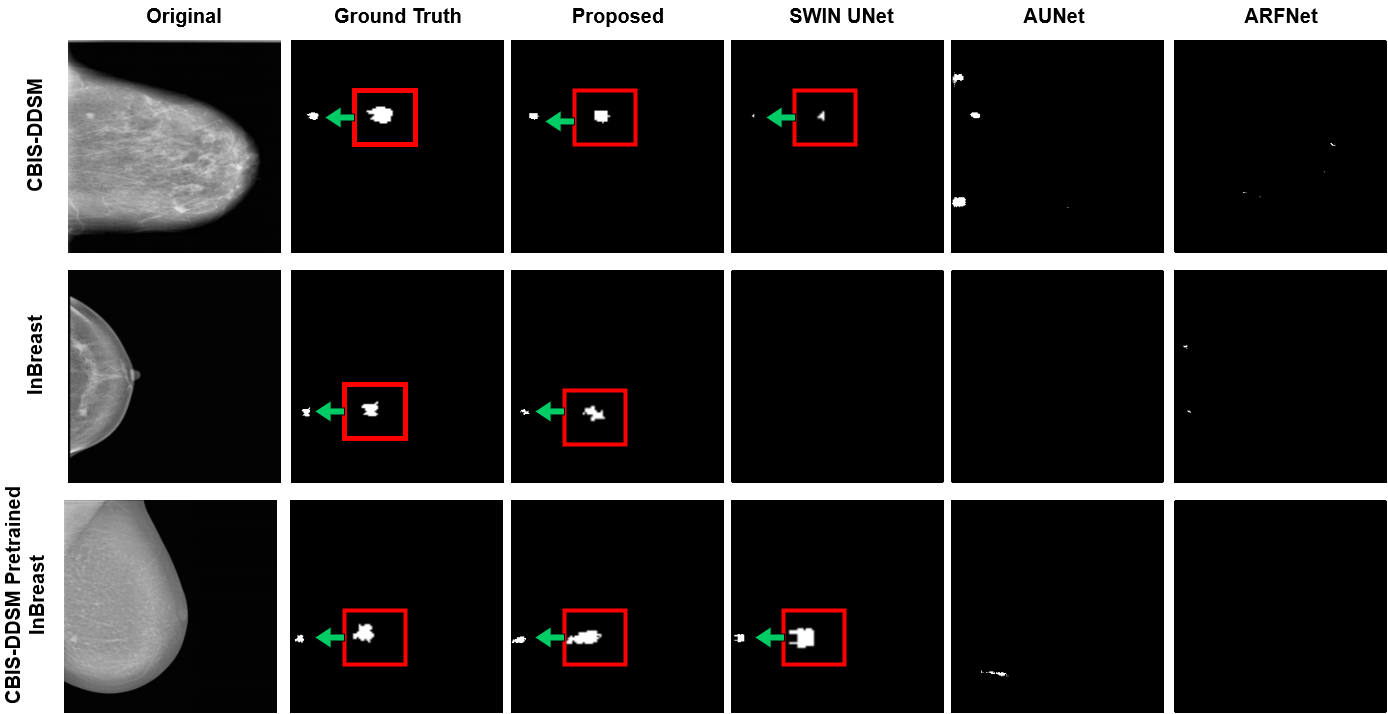}
    \caption{Qualitative performance of Swin-SFTNet vs. other architectures on three datasets. ({\color{red} Red box}) are zoomed-in images.}
    \label{fig3}
\end{figure*}
\section{Experiments}
\label{sec:experiment}

\subsection{Dataset}
\label{sec:dataset}
We evaluated our model with three publicly available datasets. We used CBIS-DDSM \cite{lee2017curated} and InBreast \cite{moreira2012inbreast}, two whole mammography segmentation datasets. All images are resized to $256\times 256$ dimension using bilinear interpolation, and the masks are resized to the same size using the nearest-neighbor technique. Both dataset contains craniocaudal (CC) and mediolateral oblique (MLO) views of breasts. From the CBIS-DDSM dataset, we separate 849 training and 69 test images based on the subtlety of 4 and 5. The masses on the test images are less than 200 pixels in size, which is 0.3\% of the whole image. The subtlety defines the visual challenge to annotate the masses for the clinician, with 1-5 grading where 1= ungradable and 5=most gradable. We use OpenCV’s contour-based technique to remove artifacts, and for enhancement, we use CLAHE. The Inbreast dataset contains 107 images, which we split into 90 training and 17 test images. The test images are separated based on any mass being less than 100 pixels or smaller.

\subsection{Hyper-parameter}
\label{sec:hparameter}
We chose $ \lambda_{bce} =0.4$, $ \lambda_{dice} =0.6$ and $ \lambda_{emb} =0.01$ (Eq.~\ref{eq6}). For optimizer, we used Adam with a learning rate of $\alpha=0.0001$, $\beta_1=0.9$ and $\beta_2=0.999$. We used Tensorflow 2.8 to train the model in mini-batches with the batch size, $b=8$ in 100 epochs which took around 1 hour to train on NVIDIA A30 GPU. The inference time is 41 millisecond per image.

\subsection{Quantitative Evaluation}
\label{sec:qevaluation}
For micro-mass segmentation tasks, we compare our model with three state-of-the-art architectures, AUNet \cite{sun2020aunet}, ARF-Net \cite{xu2022arf}, and SWIN-UNet \cite{cao2021swin} for CBIS-DDSM and InBreast, as given in Table.~\ref{table1}. AUNet utilizes attention-guided dense upsampling to retain important spatial features lost due to bilinear up-sampling. In contrast, ARF-Net uses a Selective Receptive Filed Module (SRFM) module to fuse multi-scale and multi-receptive field information and the current state-of-the-art for both breast segmentation datasets. Finally, Swin-UNet combines swin transformer blocks presented in \cite{liu2021swin} with a UNet-like structure to reach high precision in multiple organ segmentation. ARFNet and AUNet show high-performance gains against previous approaches. However, the prediction is skewed because the test set contains images with more large masses and few micro-masses. We designed the experiment to emphasize micro-mass segmentation, so we sorted the images based on the tumor size. For InBreast, we chose the last portion for testing (small than 100 px), which has the smallest breast masses. And for CBIS-DDSM, we discarded the leftover large mass test images (larger than 200 px), as we had separate training images.

For metrics, we use Dice score $=\frac{2 \times TP}{2 \times TP + FP+ FN}$,  Mean IOU (mIOU) $=\frac{TP}{TP + FP+ FN}$, Sensitivity (SEN) $=\frac{TP}{TP+FN}$, and Specificity (SPE) $=\frac{TN}{TN+FP}$. We can see from Table.~\ref{table1}, our model achieves the best score compared to others for all the metrics. We reach higher segmentation dice score over the state-of-the-art by 3.10\% on CBIS-DDSM, 3.81\% on InBreast, and 3.13\% on CBIS pre-trained model tested on InBreast ({\color{red}Given in Red}). Moreover, in qualitative comparison in Fig.~\ref{fig3}, our model can segment harder and smaller masses than other architectures. 

We also did ablation study for the embedding loss for two datasets, which are provided in Table.~\ref{table2}. With the novel loss function we have 3.14\%, 6.33\%, and 0.86\% gain for CBIS-DDSM, InBreast, and CBIS pre-trained model consecutively.   

\begin{table}[!tp]
\caption{Comparison for \textbf{CBIS-DDSM}, \textbf{InBreast} and \textbf{CBIS-DDSM Pretrained InBreast}}
\centering
\begin{adjustbox}{width=0.9\columnwidth}
\begin{tabular}{|c|c|c|c|c|c|}
\hline
\textbf{Dataset}                                                                                    & \textbf{Model}     & \textbf{Dice(\%)} & \textbf{mIoU(\%)} & \textbf{SEN(\%)} & \textbf{SPE(\%)} \\ \hline
\multirow{4}{*}{CBIS-DDSM}                                                                 & AUNet     & 14.20    & 9.14     & 29.81   & 99.40   \\ \cline{2-6} 
                                                                                           & ARFNet    & 2.54     & 1.35     & 1.44    & 99.98   \\ \cline{2-6} 
                                                                                           & SWIN UNet & 21.03    & 15.53    & 31.36   & 99.70   \\ \cline{2-6} 
                                                                                           & Proposed  & {\color{red}\textbf{24.13}}    & {\color{red}\textbf{17.44}}    & {\color{red}\textbf{33.31}}   & {\color{red}\textbf{99.72}}   \\ \hline
\multirow{4}{*}{InBreast}                                                                  & AUNet     & 12.11    & 9.35     & 15.28   & 99.94   \\ \cline{2-6} 
                                                                                           & ARFNet    & 13.95    & 8.69     & 21.21   & 99.68   \\ \cline{2-6} 
                                                                                           & SWIN UNet & 14.12    & 9.46     & 28.17   & 99.55   \\ \cline{2-6} 
                                                                                           & Proposed  & {\color{red}\textbf{17.93}}    & {\color{red}\textbf{13.10}}    & {\color{red}\textbf{20.56}}   & {\color{red}\textbf{99.81}}   \\ \hline
\multirow{4}{*}{\begin{tabular}[c]{@{}c@{}}CBIS-DDSM\\ Pretrained\\ InBreast\end{tabular}} & AUNet     & 17.24    & 12.41    & 23.49   & 99.74   \\ \cline{2-6} 
                                                                                           & ARFNet    & 2.84     & 1.53     & 65.04   & 86.30   \\ \cline{2-6} 
                                                                                           & SWIN UNet & 20.25    & 14.27    & 29.46   & 99.58   \\ \cline{2-6} 
                                                                                           & Proposed  & {\color{red}\textbf{23.38}}    & {\color{red}\textbf{17.40}}    & {\color{red}\textbf{34.54}}   & {\color{red}\textbf{99.87}}   \\ \hline
\end{tabular}
\end{adjustbox}
\label{table1}
\end{table}

\begin{table}[!ht]
\caption{Dice score  for with or w/o Feature Matching Loss}
\centering
\begin{adjustbox}{width=0.65\columnwidth}
\begin{tabular}{|c|c|c|c|}
\hline
\begin{tabular}[c]{@{}c@{}}Feature\\ Matching\\ Loss\end{tabular} & CBIS-DDSM & InBreast & \begin{tabular}[c]{@{}c@{}}CBIS-DDSM\\ Pretrained\\ InBreast\end{tabular} \\ \hline
With                                                              & \textbf{24.13}     & \textbf{17.93}    & \textbf{23.38}                                                                     \\ \hline
Without                                                           & 20.72     & 11.60    & 22.52                                                                     \\ \hline
\end{tabular}
\end{adjustbox}
\label{table2}
\end{table}

\section{Conclusion}
\label{sec:conclusion}
In this paper, we proposed Swin-SFTNet, with a novel Spatial Feature Expansion and Aggregation Block (SFEA) block, which captures the global context of the images and fuses it with the local patch-wise features. Moreover, we also integrate a novel embedding loss that computes the similarities between the encoder and decoder block's patch-level features. Our model outperforms other architectures in micro-mass segmentation tasks in two popular datasets.

\section{ACKNOWLEDGEMENTS}
No funding was received for conducting this study. The authors have no relevant financial or non-financial interests to disclose.

\section{COMPLIANCE WITH ETHICAL STANDARDS}
This research study was conducted retrospectively using human subject data made available in open access by \cite{lee2017curated,moreira2012inbreast}. Ethical approval was not required as confirmed by the license attached with the open access data.

\bibliographystyle{IEEEbib}
\bibliography{strings,refs}

\end{document}